\documentclass[a4paper,12pt]{article}
\usepackage{jheppub}
\usepackage{amsmath,amssymb,multirow,array}

\title{{\bf The Energy Loss of a Heavy Quark Moving in a Viscous Fluid}}

\author{Navid Abbasi$^a$\footnote{Abbasi@ipm.ir}, \ Ali Davody$^b$\footnote{Davody@ipm.ir},\\
\small{\emph{$^{a}$Department of Physics, Sharif University of Technology,}} \\
\small{\emph{P.O. Box 11365-9161, Tehran, Iran}} \\ [1mm]
\small{\emph{$^{b}$ School of Particles and Accelerators, Institute for Research in Fundamental Sciences (IPM), P.O. Box 19395-5531, Tehran, Iran}} 
}

\abstract{
To study the rate of energy and momentum loss of a heavy
 quark in QGP, specifically in the  hydrodynamic regime, 
we use fluid/gravity duality  and construct  a perturbative
 procedure  to find the string solution in  gravity side.
 We show that by this  construction the drag force exerted on the quark
 can be computed  perturbatively, order by order in a
boundary derivative expansion.
At   ideal order, our result is just the drag
force exerted on a  moving quark in thermal plasma with thermodynamics variables promoted to become
local functions of space and time. Furthermore, we apply this procedure
to a transverse quark in Bjorken flow  and compute the  first-derivative corrections, 
namely the viscous corrections, to the drag force. 
}

\begin{document}

\setlength{\baselineskip}{16pt}
\begin{titlepage}
\maketitle

\vspace{-36pt}

\thispagestyle{empty}
\setcounter{page}{0}
\end{titlepage}

\renewcommand{\baselinestretch}{1}  
\tableofcontents
\renewcommand{\baselinestretch}{1.2}  

\section{Introduction}
\label{1}
Studying systems out of equilibrium is  an interesting subject in physics.
Depending on how it evolves towards equilibrium,
there are different methods to study the system. In a familiar
case, when the deviation from equilibrium has small amplitude
everywhere, one can  apply linear response theory to
study the dynamics of the system. Computing retarded Green's functions in
equilibrium, we can extract certain information about  the dynamics
of the system slightly deviated from its equilibrium state.  
However, when the understudying  system is described
by a strongly coupled field theory, we are not  able to do such
computations through perturbative methods. In such cases,  AdS/CFT
correspondence  appears as a powerful tool.

AdS/CFT conjecture in its original statement, relates the type IIB
string theory on $\mathrm{AdS}_{5}\times S^5$ space-time to
the four-dimensional $\mathcal{N}=4$ SYM gauge theory \cite{Maldacena:1997re}.
On the other hand, there are many examples which have been studied with the
holographic description of AdS/CFT in which, a strongly coupled
field theory living on the boundary of the AdS space is pictured
to the weakly coupled gravity theory in the bulk of AdS \cite{Witten:1998qj}.
Accordingly, this holographic picture,  provides a tool to study the
strongly coupled systems out of equilibrium \cite{Hubeny:2010ry}.

Although our familiar strongly coupled systems do not exactly
coincide with the  $\mathcal{N}=4$ SYM gauge theory, we apply this theory
as a  useful model  to study such  systems through  AdS/CFT
correspondence. This modelling  shows agreement with real physical
systems in particular circumstances. For example in QGP 
experiments at RHIC or at LHC, a gas of quarks and gluons is
produced with the temperature about 170 MeV  which is strongly coupled.
Although QCD as the underlying theory in these experiment 
is distinct from $\mathcal{N}=4$ SYM gauge theory, but this has not  
prevented us to  apply this to to study of QCD in QGP experiments.
Specially in the hydrodynamic regime, computation of the
transport coefficients is known as  a remarkable success of AdS/CFT 
\cite{Kovtun:2004de} providing results in agreement with experiment.

  One of the other significant subjects  from both experimental
  and  theoretical standpoints, is the motion of quarks created during
  heavy-ion collisions in the plasma. Up to now, invoking AdS/CFT, 
  different mechanisms have been introduced to describe the rate of 
  energy and momentum loss  through this motion;  \cite{Herzog:2006gh,Gubser:2006bz,CasalderreySolana:2009ch,Fadafan:2008bq,Chernicoff:2008sa}. 
  There is a simple holographic picture dual to these problems. The  probe
  quark moving  in plasma is mapped to a probe string  in the AdS space. So instead
  of studying  quark's motion in a strongly coupled system
  one can simply trace the classical dynamics of a string in
  gravity side. From now on, we  concentrate on the drag force formalism \cite{Herzog:2006gh,
  Gubser:2006bz} to compute the energy loss in this paper. 
  A common part in both  \cite{Herzog:2006gh,
  Gubser:2006bz} is that the   authors have studied  the uniform motion 
  of a quark in the boundary plasma. This quark is  the end point of a string stretching
  in the AdS bulk. So in order to move uniformly, it should be forced
  by an external source, i.e. an electric field. The momentum rate
  flowing down to the string has been interpreted as drag force
  exerted from the plasma on the quark. It has been shown in \cite
  {Herzog:2006gh,Gubser:2006bz} that when a quark is dragged with a
   constant velocity $v$ in a thermal plasma at 
   temperature $T$, the rate of the energy and momentum  loss of the
   quark in the plasma frame are given by
    \begin{equation}
  \frac{dp}{dt}=\frac{1}{v}\;\frac{dE}{dt}=-\frac{\pi}{2\alpha'}\; T^2\;\frac{v}{\sqrt{1-v^2}}. \label{1}
    \end{equation}
    In these derivations, besed on AdS/CFT
     correspondence, the  gravity set up dual to the  thermal field
      theory has been taken as a uniform 5-dimensional AdS black
      brane with the metric
    \begin{equation}
      \mathrm{d}s^2=G_{MN} \mathrm{d} x^{M} \mathrm{d}x^{N} =\frac{ \mathrm{d}r^2}{r^2 f(br)} +r^2(P_{\mu\nu} -f(br)u_{\mu} u_{\nu}) \mathrm{d} x^{\mu} \mathrm{d}x^{\nu} \label{1'}
  \end{equation}
where $M=(r,\mu)$, $f(r)=1-\frac{1}{r^4}$ and the temperature of the black brane
 is given by $T=1/\pi b$. Also  $P_{\mu \nu}=u_{\mu}u_{\nu}+\eta_{\mu\nu}$
  is a projection tensor where
$u^{\mu}$ is the four-velocity of the
 plasma. For a plasma at rest in  the laboratory frame $u^{\mu}=(1,0,0,0)$.
 
   From the experimental  viewpoint, an important case
  to study is a moving quark in an expanding plasma  where, local
  thermal equilibrium has been attained. It can be generated
  in the time interval between thermalization and hadronization,
  after  scattering of two massive nuclei \cite{Kolb:2003dz}.
  In this  time interval, the excited matter is expanding and
    cooling down such that at any point, hydrodynamic description
    is valid and  therefore, dynamics of the expansion can be
    described by the corresponding equations of motion \cite{Banerjee:2011tg}.
    Beforehand  AdS/CFT, there was not any theoretical approach to
    study the hydrodynamic regime in strongly coupled field theories.
    In fact,  AdS/CFT  has opened a new window to research:
    study of CFT hydrodynamic from gravity. First efforts in this context
    were made by \cite{Policastro:2002se,Policastro:2002tn} which were
    confirmation of existence of the hydrodynamic regime in a  $\mathcal{N}
    =4$ SYM gauge theory. In another remarkable development, Janik
    and Peschanski   found an asymptotic gravity dual to the hydrodynamical flow
    in QGP, namely Bjorken flow,  from a nonlinear derivation \cite{Janik:2005zt}.
    In 2007, BHMR introduced  the full version of  fluid/gravity
    duality   \cite{Bhattacharyya:2008jc} in which, they showed  that
    there is a one to one map between fluid dynamical flows of a CFT on the
    boundary with the long wavelength perturbations of the  AdS black brane in
    the bulk.

    It may be interesting to investigate the drag force exerted on a quark  in a fluid 
    dynamical flow as a 
    phenomenological problem related to QGP experiments. In this direction,
    the drag force exerted on a quark moving through Bjorken flow was first computed
    in  \cite{Giecold:2009wi} and later in \cite{Stoffers:2011fx}. In these works,
    Bjorken fluid has been considered as an ideal fluid. To be more  precise,
    one can go further in the hydrodynamic expansion and study the viscous effects.
    For instance, the gravity set up introduced in \cite{Heller:2008fg},
    may be exciting to such investigation in Bjorken fluid.  Although the perturbative method
    given in  \cite{Heller:2008fg} is clear, perturbatively solving the string in this background 
    seems so hard.

    As discussed above, fluid/gravity correspondence maps  each long wavelength
    perturbation  around an AdS black brane to  a hydrodynamical flow on the
    boundary of AdS.
      Apart from sketching the mentioned map, a key point in
    fluid/gravity duality is that it provides a perturbative technology
    to find the regular solutions in AdS space. Similarly,  based on fluid/gravity
    duality, we construct a perturbative procedure to find the classical string 
    solution  in the long wavelength perturbed  AdS background.  The choice of
    coordinates has a key role in constructing our perturbative procedure. We 
    exploit the   Eddington-Finkelstein coordinates in which the extension
    of the string in the boundary directions is finite. It is in contrast to the
    coordinates systems taken in \cite{Gubser:2006bz,Herzog:2006gh} where, the 
    embedding of the string  has an infinite extension in the boundary spatial 
    directions. This delicate point is one of our motivations to construct  the  perturbative
    procedure given introduced in this paper.

    It is important to note that constructing such perurbative procedure is the key 
    issue in development of our idea in this work. Using this procedure, one can compute  certain  physical quantities
    related to a quark in fluid medium perturbatively. For example  we proceed to
    compute the drag force exerted on a quark moving in a fluid dynamical flow.
    As a main result, we show that at ideal order, drag force can be computed in a 
    flow-independent manner. 
     In addition, we extend  this procedure  to higher orders in 
    perturbation. We discuss it will be possible to compute corrections to the drag
     force  perturbatively, order by order in a boundary
    derivative expansion. Finally, we apply this method to a transverse quark created
    in Bjorken fluid and compute the drag force perturbatively up to first order.
    It would be interesting to use our formula and modify the result of \cite{Gubser:2006qh}
     where compared to the result of \cite{vanHees:2004gq}.

    This paper is organized as follows. We begin the \eqref{2} with a brief
    review of drag force derivation in String/CFT duality in a world-sheet covariant
    approach. In \eqref{three} we apply the statement of \eqref{2} to compute the space-time
    covariant drag force exerted on a quark in a global boosted plasma in. \eqref{4}
     is devoted to the core  of this paper in which after reviewing some key points in
     fluid/gravity correspondence, we explain our perturbative procedure. Then, we apply
     this method to find  classical string solution in  gravity background dual to a
     boundary fluid dynamical flow. As a preliminary application of our perturbation method, we
     study a quark moving in a slowly time varying plasma in \eqref{5}. In  \eqref{six} we apply
     our perturbation to Bjorken  fluid and compute first-derivative correction to the drag
     force exerted on a  transverse  quark in this fluid. Finally in \eqref{7} we end with a discussion of results
    and of possible future projects.

\section{World-sheet covariant description of drag force}
\label{2}

Consider a quark moving with velocity$u_{q}$ in the  boundary field theory.
 According to the string/CFT duality, such quark  is the end point of an open string,
 embedded in a one higher dimensional space-time dual to the boundary field theory.
 The dynamics of the classical string is described by the  Nambu-Goto action as follows
 \begin{equation}
          S=-\frac{1}{2\pi\alpha'}\int\mathrm{d}\sigma\;\mathrm{d}\tau\;
           e^{\frac{\Phi}{2}}\sqrt{-g},\;\;\;\;\sigma\in[0,\sigma_{1}] \label{kk}
          \end{equation}
 where $\Phi$ is the background dilaton field. Also in the above
 expression, $g=detg_{\alpha\beta}$ where  $g_{\alpha\beta}=G_{\mu\nu}\;
 \partial_{\alpha}x^{\mu}\partial_{\beta}x^{\nu}$. Free motion of the
quark in the plasma is dissipative.
 AdS/CFT duality ascribes  this dissipation effect to  flowing
 the  momentum down to the string in the bulk.  Therefore,
  to have a uniform motion, the quark should be forced by an
  external source such as an electric field on the boundary.
  In such a case, the mentioned flow rate is interpreted as
  the drag force exerted on the quark.

  By construction, drag force should is  a vector in space-time and a scalar
  on the world-sheet. In order to  have it in a world-sheet covariant form,
  consider the world-sheet currents as following
   \begin{equation}
   \Pi_{M}^{\alpha}=\frac{\partial{\cal{L}}}{\partial({\partial_{\alpha}X^{M}})}=-\frac{e^{\frac{1}{2}\Phi(x^{\mu})}}{2\pi\alpha'}\;\;\sqrt{-g}\;\;P_{M}^{\alpha} \label{3'}
   \end{equation}
 where $\alpha$ and $M=r,\mu$ describe the world-sheet and space-time
 directions respectively. In contrast to $P_{M}^{\alpha}$, these
 currents are not  the world-sheet vectors. According to the previous paragraph
  discussion, to compute  drag force we should find ingoing flux of
  the energy and momentum across the time-like boundary of the world-sheet, 
  the $\sigma=\sigma_{1}$ curve. Assuming $n^{\alpha}$ as the inward normal vector
    to  this curve, the  covariant 4-drag force is given by:
   \begin{equation}
   F_{\mu}=\frac{e^{\frac{1}{2}\Phi(x^{\mu})}}{2\pi\alpha'}P_{\mu}^{\alpha }\;n_{\alpha}.  \label{3''}
      \end{equation}
It is clear that the choice of the gauge determines what combination
of $P_{M}^{\alpha }$s is equal to  the drag force. Let us recall that 
the relativistic four-force, $F^{\mu}=\frac{d}{d\tau}p^{\mu}$, 
 exerted on the quark  should be perpendicular to the quark four-velocity:
\begin{equation}
F_{\mu}\;u_{q}^{\mu}=0.  \label{3'''}
\end{equation}
 For a quark moving with velocity $u_{q}$
 in the  1-direction of the plasma frame, applying \eqref{3'''} to
   $F_{\mu}=\gamma\;(\frac{dE}{dt},\frac{dp}{dt},0,0)$ ensures that
\begin{equation}
\frac{dp}{dt}=\frac{1}{u_{q}}\frac{dE}{dt}  \label{3"}
\end{equation}
which can be  obviously observed in \eqref{1}  and all of our
 perturbative results in next sections.

\section{Moving quark in a boosted thermal plasma}
\label{three}
 Transforming the coordinates in AdS black brane metric \eqref{1'} to
 Eddington-Finkelstein coordinates, we find the
 metric of  gravity space-time dual to the boundary  boosted
 thermal plasma:
 \begin{equation}
 \mathrm{d}s^2=G_{MN} \mathrm{d} x^{M} \mathrm{d}x^{N} =-2u_{\mu} \mathrm{d}x^{\mu} \mathrm{d}r+r^2 \left(P_{\mu\nu} -f(br)u_{\mu} u_{\nu} \right) \mathrm{d} x^{\mu} \mathrm{d}x^{\nu}  \label{4'}
\end{equation}
 where, the  boosted plasma with velocity $u^\mu$ and temperature
  $T=\frac{1}{\pi b}$ lives on the  boundary of space-time,
  $r\to\infty$. It is significant to note that \eqref{4'}
  introduces an exact $4+1$ parameters family of Einstein
   equations solutions \cite{Rangamani:2009xk}.

 To study  the energy and momentum loss of  a moving quark in this
  system, we consider the  four-velocity of plasma
  to be  $u^\mu_{f}=(u^0,u^1,0,0)$ and for simplicity assume that
   the quark is moving in the $1$-direction with a constant velocity
    $u_q$. Above-mentioned  quark is the  end point of an open string embedded
     in a 5-dimensional space-time by the metric \eqref{4'}. Dynamics of the string
      is governed by the Nambu-Goto action (\ref{kk}) where
       dilaton factor can be dropped in  the above background.
        In  the static gauge $(\sigma,\tau)=(r,t)$, embedding of
        the string  is given by $X^{M}(r,t)=(r,t,X^{i}(r,t))$.
        Plugging
\begin{equation}
X^{1}=u_{q}t+\xi(r),\;\;\;X^{2}=X^{3}=0 \label{2' }
\end{equation}
  in (\ref{kk}), we reach to following expression:
    \begin{equation}
    {\cal{L}} = \sqrt{A\xi'^{2}+ {2}B\xi'+C} \label{kkk}
    \end{equation}
    where $A$, $B$ and $C$ are some constant coefficients  which depend
     on the bulk metric elements and the velocity of the quark (\ref{111}).
     As it can be observed, $\xi$ enters into the world-sheet  lagrangian
     only through its derivative $\xi'$. So
      instead of solving EOM, one can exploit the  conservation
       equation of  the conjugate momentum of $\xi$:
     $ \pi_{\xi}=\partial {\cal{L}}/\partial \xi'$.
       Deriving   $\xi'$  in terms of $\pi_{\xi}$, we find
  \begin{equation}
  \xi'=-\frac{B}{A}\pm\frac{1}{A}\sqrt{\pi_\xi^2\frac{B^2-AC}{\pi_{\xi}^2-A}}. \label{aa}
\end{equation}
    It can be simply checked that ${B^2-AC}$ vanishes at some point in the bulk.
    Obviously, for the second term to be real and so the induced metric
    to be non-degenerate,
    numerator and denominator should have common root, namely $r^*$.
    Solving $B^2-AC=0$, $r^*$ is obtained as
      \begin{equation}
      r^*=\frac{1}{b}\;\sqrt{\frac{1-\beta u_{q}}{\sqrt{1-\beta^2}\sqrt{1-u_{q}^2}}}. \label{12}
      \end{equation}
  Demanding the denominator to vanish at  $r^*$, determines the  value of $\pi_{\xi}$  as:
    \begin{equation}
     \pi_{\xi}=\sqrt{A(r^*)}=   \frac{1}{b^2}\frac{u_{q}-\beta}{\sqrt{1-u_{q}^2}\sqrt{1-\beta ^2}} \label{ee}
    \end{equation}
    where $\beta=\frac{u^{1}}{\sqrt{1+{u^{1}}^2}}$ is the 1-component
     of the plasma  velocity in the boundary lab frame (LF).
     Inserting (\ref{ee}) in  (\ref{aa}), we have:
    \begin{equation}
     \xi'=\frac{u_{q}-\beta}{r^2 f(br)\sqrt{1-\beta^2}}(-1\pm\frac{1}{b^2 r^2}), \label{cc}
    \end{equation}
where  the choice of $+$ corresponds to flowing energy and momentum from the quark
into the bulk.

\subsection{Drag force computation}
\label{3.1}

 Let us recall that our  gauge choice,
  fixes the inward normal vector to the boundary of the world-sheet at
  $r \to \infty$ as $n_{\alpha}=(-1,0)$. So the $1$-component of
   4-drag force may be written as:
 \begin{equation}
 F^{1}=F_{1}=P^{r}_{1}=\left(\frac{1}{\sqrt{-g}}\Pi^{r}_{1}\right)_{r\to \infty} \label{7'}
 \end{equation}
    where
    \begin{equation}
    \Pi_{1}^r=-\frac{1}{2\pi\alpha'} G_{1\nu} \frac{(\dot{X}.X)(\dot{X}^\nu)'-(\dot{X})^2(X^{\nu})'}{\sqrt{-g}}=-\frac{1}{2\pi\alpha'}\pi_{\xi}. \label{8}
\end{equation}
     Translational invariance of (\ref{kkk}) ensures the conservation of the world-sheet currents,
    $\partial_{\alpha}\Pi_{\mu}^\alpha=0$.
    Knowing the value of $\Pi_{1}^{r}$ at $r^*$,  one can integrate
    this equation to compute $\Pi_{1}^{r}$ as a function of radial
    coordinate. Since the space-time
     is static, it turns out that $\Pi_{1}^{r}$ is constant along
     the string and equal to $\Pi_{1}^{r}(r^*)$. Therefore we obtain
     the 1-component of 4-drag, $F^{\mu}=(\gamma \frac{dE}{dt}, \gamma \frac{dp}{dt},0,0)$, as follows:
    \begin{equation}
    F^{1}=\frac{1}{\sqrt{1-u_{q}^2}} \;\frac{dp}{dt}=-\frac{1}{2\pi \alpha'}\frac{1}{\sqrt{1-u_{q}^2}}\pi_{\xi}   \label{8'}
    \end{equation}
    where $dp/dt$ is the 1-component of 3-drag force in  LF.

It can be simply seen  that for $u_{q}>\beta$, drag force
 is pointed to  opposite direction of the quark motion while for
  $u_{q}<\beta$ aligns in the quark  direction of motion. Expectedly,
   when the quark is moving  with the velocity equal to that of the
   plasma, does not sense any drag force.

  To compute the rate of energy loss, we should work out $\Pi_{1}^{t}$. 
 After a simple  calculation we have
 \begin{equation}
F^{t}=\frac{1}{\sqrt{1-u_{q}^2}}  \;  \frac{dE}{dt}= -\frac{1}{2\pi \alpha'}\frac{u_{q}}{\sqrt{1-u_{q}^2}}\pi_{\xi} .  \label{8''}
\end{equation}
   It is interesting to note that for $\beta=0$, the rate of  energy and
   momentum loss in the LF obtained from \eqref{8'} and \eqref{8''}
   exactly coincide with \eqref{1}.  It is expectable
   that one could obtain the above expression by a lorentz transformation
   of \eqref{1}.
   In addition, \eqref{8'} and \eqref{8''} remind us that instead
   of studying string picture dual to a moving quark in  global
   thermal plasma, one can apply string/CFT duality to  a rest
   quark in a global boosted thermal plasma, since both
   $(u_{q}=v,\beta=0)$ and $(u_{q}=0,\beta=-v)$ cases have common
   results from (\ref{ee}).

\subsection{Embedding of the string}
\label{3.2}
\begin{figure}
\centering
\includegraphics[scale=.3]{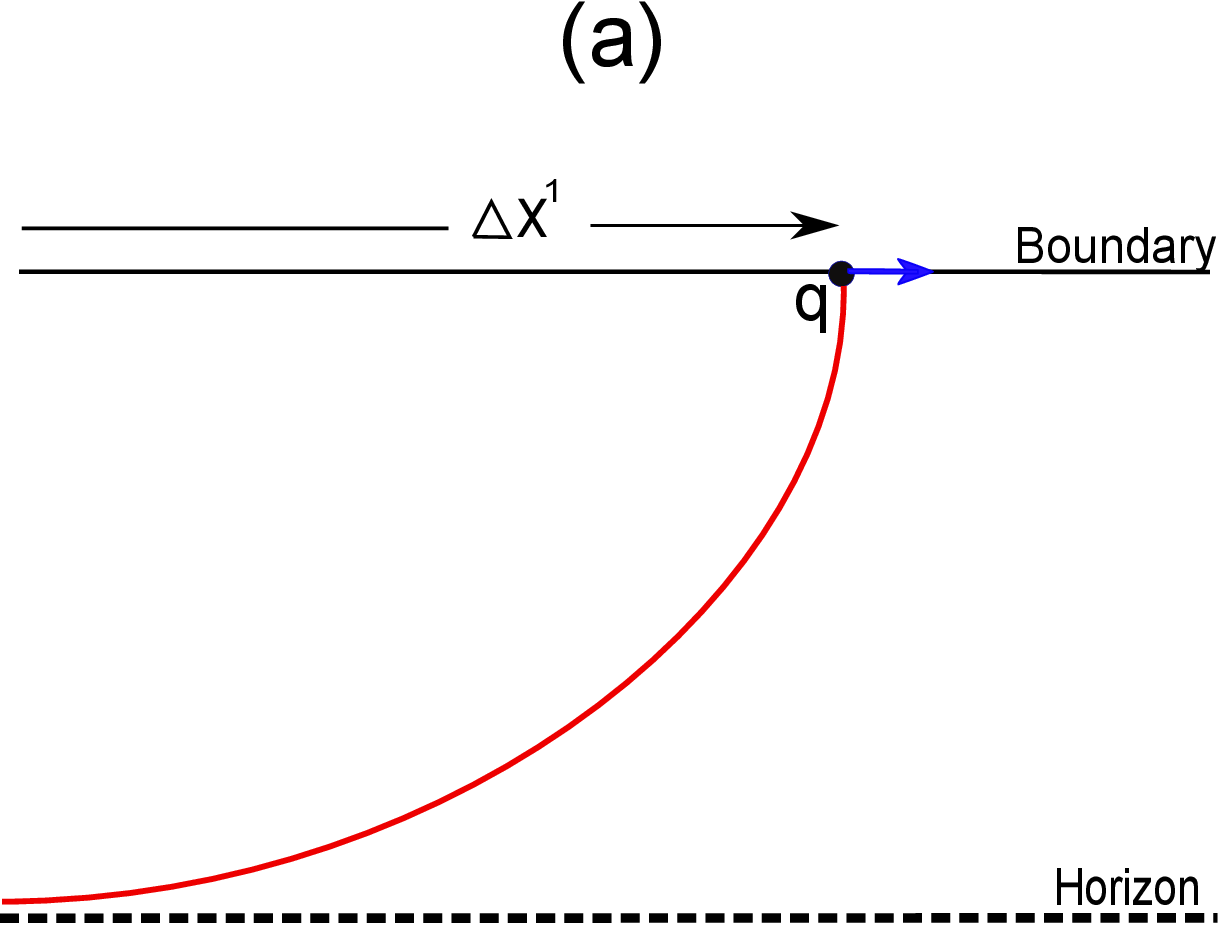}
\hspace{2cm}
\includegraphics[scale=.3]{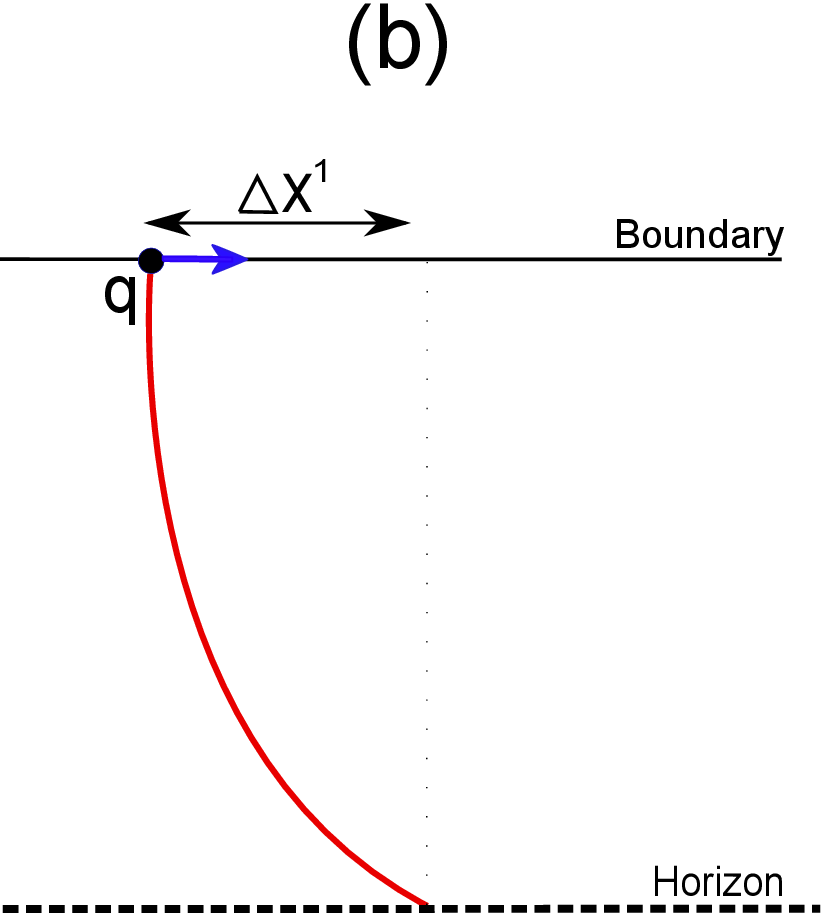}
\caption{Comparison of the string embedding in the bulk 
between two different coordinates systems. In the Eddington-Finkelstein
coordinates, string has been stretched in front of the moving quark in 
a finite range on the boundary.}    \label{fig:1}
\end{figure}

   The  embedding of  the string in the bulk space-time is the
    solution of equation (\ref{cc}). Assuming $u_{q}>\beta$  and
     picking up the plus sign to ensure  the momentum flowing 
     down the string, we have
    \begin{equation}
    \xi(r)=-\frac{u_{q}-\beta}{\sqrt{1-\beta^2}}\;b\;\left(\arctan(rb)-\frac{\pi}{2}\right) \label{9}
    \end{equation}
    where we have chosen  the constant of integration such that $X^{1}$
   to  be identified by $x=vt$ on the boundary. In Fig.\ref{fig:1} the embedding
   of the string has been compared in two different coordinates systems. 
   Figure (a) shows  the profile of the string for $u_{q}>\beta$ in a
   constant time slice  of the metric \eqref{1'}.  In case (b) we have shown
   the same string has been considered in the coordinates system
   given in  \eqref{4'}. Although the quark is moving to  the
   right, the string trails out in front of the quark in this case.
   It is not so surprising because, the time coordinate in
   \eqref{4'} is the ingoing Eddington-Finkelstein time in which
   all of events on an ingoing null geodesic are simultaneous.

   In contrast to the former case in which the extension of the string
   in the 1-direction is infinite, in the latter case, the string
   has occupied a finite range in the 1-direction. Substituting the
   value of the horizon radius in \eqref{9} gives the length of
   this finite extension in the LF as following:
    \begin{equation}
    \Delta x^{1}=\frac{\pi}{4}\;b\;\frac{u_{q}-\beta}{\sqrt{1-\beta^2}}. \label{11}
\end{equation}

       The finiteness of $\Delta x^{1}$, as a direct consequence of
       applying the Eddington-Finkelstein coordinates will play a key role in study of  moving quark
         in  fluid medium in  next sections.
      
\subsection{Covariant results}
\label{3.3}

 In this subsection we want to present our previous result in the frame-independent 
 expressions.   It should be noticed  that we have only two independent four-vectors in
 our analysis, four-velocity of the quark and of the plasma. So to have a covariant drag force
we construct  a linear combination of these velocities and demand the perpendicularity condition
in  \eqref{3'''}. We obtain
    \begin{equation}
    F^{\mu}=\frac{1}{2\pi \alpha'}\frac{1}{b^2}((u_{q}.u_{p})u^{\mu}_{q}+u^{\mu}_{p}) \label{8'''}
\end{equation}
where $u^{\mu}_{q}$ and $u^{\mu}_{p}$  are the four-velocity
of the quark and plasma respectively. The constant factor has
been fixed by going to the LF where  $u^{\mu}_{q}=\gamma_{q}(1,u_{q},0,0)$,
 $u^{\mu}_{p}=(1,0,0,0)$, $ F^{\mu}=\gamma_{q}(\frac{dE}{dt},\frac{dp}{dt},0,0)$,
  and applying the LF results given in \eqref{1}.

The position of the world-sheet horizon, $r^*$, as a lorentz
scalar can be given in following covariant expression:
\begin{equation}
r^*=\frac{1}{b}\sqrt{-u_{q}.u_{p}}. \label{8"}
\end{equation}
\section{Gravity dual to a quark moving through a fluid flow}
\label{4}
       Our final goal in two next sections is  to investigate the uniform
       motion of a  quark in a conformal fluid described by the hydrodynamic
        regime of an interacting field theory. According to the string/CFT
         picture, to study the quark motion in such an interacting field
          theory, the string should be solved  in a bulk gravity background
           dual to the boundary fluid flow with appropriate boundary
            conditions. So it is  needed to recall some of the main
            results of fluid/gravity correspondence.

\subsection{Review of the fluid/gravity duality}
\label{4.1}
        Consider a 4-dimensional field theory living on the boundary of
         an asymptotically $AdS_{5}$  space-time. It has been shown that
         there is a map between the  hydrodynamic regime of the strongly
          coupled boundary field theory and a class of inhomogeneous,
          dynamical black hole solutions in AdS space-time \cite{Rangamani:2009xk,Hubeny:2011hd}. In fact
           the long wavelength perturbations of an asymptotically AdS
            black brane are mapped to corresponding dynamical flows of
             a conformal fluid on the boundary.

    According to \cite{Bhattacharyya:2008jc}, the dual gravity of an arbitrary boundary flow can be
    found perturbativly, order by order in a boundary derivative expansion.
     It means that the  gravity dual to a boundary fluid flow with velocity
     $u^{\mu}(x^{\alpha})$ and temperature $T(x^{\alpha})$ can be obtained
     perturbatively, order by order by adding appropriate correction terms
       to the localized version of the metric \eqref{4'};
  \begin{equation}
  \mathrm{d}s^2=-2u_{\mu}(x^{\alpha}) \mathrm{d}x^{\mu} \mathrm{d}r+r^2 [P_{\mu\nu}(x^{\alpha}) -f(b(x^{\alpha})r)u_{\mu}(x^{\alpha}) u_{\nu}(x^{\alpha})] \mathrm{d} x^{\mu} \mathrm{d}x^{\nu}+\mathrm{d}s^2_{cor}, \label{13}
 \end{equation}
  where $P_{\mu\nu}=\eta_{\mu\nu}+u_{\mu}u_{\nu}$ is the projection tensor. The first-order 
  correction of the metric has been   given in \ref{2221}.

  Although the above gravity set up seems so general, one may be interested to take into account
  forced fluid flows rather than such free flows on the boundary. This is the subject of
   \cite{Bhattacharyya:2008ji}
 in which,  the bulk solution dual to an arbitrary fluid dynamical flow of the
   boundary field theory with an arbitrary slowly varying  coupling, living on a
   weakly curved manifold,  has been found up to second order in boundary derivative
    expansion. In order to describe such dynamical flows, instead of $S=\int \cal{L}$
     the action of the boundary field theory is in the form  $S=\int \sqrt{g}e^{-\phi}\cal{L}$
      where $g_{\mu\nu}$ is the weakly curved metric of the boundary and $\phi(x^{\mu})$ is
       an arbitrary slowly varying function. From this action, the Navier-Stokes equations
       will appear with a forcing term in the right hand side:
  \begin{equation}
 \nabla_{\mu}T^{\mu\nu}=e^{-\phi}\;\cal{L}\; \nabla^{\nu}\phi. \label{14}
\end{equation}
   In \cite{Bhattacharyya:2008ji}, it has been explained that the gravity dual to an arbitrary
    general fluid dynamical flow in this field theory is a long wavelength
     solution of the Einstein-dilaton system with appropriate boundary
      conditions. Note that for  dilaton field $\Phi(r,x^{\mu})$ in
       the bulk, the boundary condition  is the necessity of  this field
        to be asymptoted to the given slowly varying function $\phi(x^{\mu})$
         on the boundary. Requiring expected boundary conditions,
         both metric of the bulk and $\Phi(x^{\mu})$ can be computed
           perturbatively, order by order in  boundary derivative
           expansions. This perturbative procedure adds derivative correction
           terms to  zero-order expressions  of the  metric and
            the dialton field in each order of perturbation:
 \begin{equation*}
    \mathrm{d}s^2=-2u_{\mu}(x^{\alpha}) \mathrm{d}x^{\mu} \mathrm{d}r+r^2 \;\;[P_{\mu\nu}(x^{\alpha}) -f(b(x^{\alpha})r)u_{\mu}(x^{\alpha}) u_{\nu}(x^{\alpha})] \mathrm{d} x^{\mu} \mathrm{d}x^{\nu}+\mathrm{d}s^2_{cor}
    \end{equation*}
    \begin{equation}
    \Phi(r,x^{\mu})=\phi(x^{\mu})+\Phi_{cor}(r,x^{\mu}) \label{155}
    \end{equation}
  where $P_{\mu\nu}=g_{\mu\nu}+u_{\mu}u_{\nu}$ is the projection tensor.
  The first-order correction of the metric  and the dilaton have been 
  given in \ref{2222}.

Emphasised by the authors of \cite{Bhattacharyya:2008jc},
 a key point in all above derivations
is the application of ingoing Eddington-Finkelstein coordinates
 which  in addition to a transparent presentation of horizon
 regularity, provide a clean   picture of  bulk gravity dual
  to the  locally equilibrated fluid  domains on the boundary.
   In this picture, each locally equilibrated domain in the
    boundary fluid,  extends along the ingoing null geodesics
    in an entire tube into the bulk. The bulk space-time within
     each tube is approximately a uniform AdS black brane where
     the width size of these tubes is the scale of variations
     in the fluid, namely $L$. As we know, having a gravity
      perturbative solution in boundary derivative expansion
      is possible, provided
 \begin{equation}
 L\;T\;\gg\;1            \label{15a}
\end{equation}
 at each point on the
      boundary. $T$ is the local temperature at the point.

       Fluid/gravity duality has a specific characteristic.
       In each order of perturbation the relevant gravity solution
       is dual to  the fluid dynamical flow  in one lower order.
       Nevertheless, the stress tensor of the fluid will be determined
       at the same order to which the metric has been solved.     
       Let us give an example. The first-order
       gravity solution is dual to an ideal fluid flow (zero order in 
       hydrodynamic expansion) on the boundary
       but it  provides the viscous fluid energy momentum  tensor 
       (first-order in hydrodynamic expansion). 
       \cite{Bhattacharyya:2008jc}.

   \subsection{Construction of the perturbative procedure}
   \label{4.2}
   After the above  brief review of fluid/gravity duality,
   we are going  to study the quark motion in a fluid background.
   Going from global equilibration state of the  plasma  to the
   hydrodynamic regime, the physical quantities related to the
   quark motion get hydrodynamic  corrections. Meanwhile,
   the dual string picture of the quark changes accordingly.
   The main idea of our paper is that for a wide range
   of  quark velocities, one can compute these corrections
   perturbatively, order by order in a boundary derivative expansion.

   The key point that allows us to use hydrodynamic perturbation procedure is
   exploiting the Eddington-Finkelstein coordinates by which, the whole string
   extension in the bulk may lie within just one tube. If such an extension
   exists, we can use the $ultralocal$ method described
   in \cite{Bhattacharyya:2008jc}to find the string solution perturbatively.
   To do that, we start with a string extended in the bulk and restricted to a tube with
   size $1/T$. At zero order of perturbation, this assumption allows us
   to neglect the velocity and temperature variations on the  the boundary
   slice of the tube, namely the boundary patch.
   Simply speaking, the difference in values of hydrodynamic variables
   between two different points in a patch, is in the
   next order of perturbation.

 So far,  we have discussed on a  string embedded in a uniform AdS
 black brane space-time. As a result the drag force can be computed just like
  \eqref{three} where, the velocity of  plasma can be chosen equal to the fluid
 velocity in an arbitrary point in the boundary patch i.e. in the position of 
 the quark.
 So, as far as we can neglect the variation of the hydrodynamic variables in a patch,
 the drag force exerted on a quark in a fluid can be computed via \eqref{8'''}
 with the fluid variables promoted to become local functions of the quark position:
   \begin{equation}
    F^{\mu}(x^{\alpha})=\frac{1}{2\pi \alpha'}\;\frac{1}{b^2(x^{\alpha})}\;\left[(u_{q}.u_{f}(x^{\alpha}))u^{\mu}_{q}+u^{\mu}_{f}(x^{\alpha})\right] \label{15''}
   \end{equation}
   where $u^{\mu}_{f}$ is the fluid four-velocity and $x^{\alpha}$ are  the coordinates of the quark position.
   Since the  local profile of the fluid has entered, th expression
   identifies  the drag force at ideal order. It is
   exactly similar to  identification of the ideal fluid stress tensor in
   \cite{Bhattacharyya:2008jc}.

    Now, we want to explore that under which circumstances in boundary side, the
    whole string extension stays in a single tube? Consider a one directional 
    fluid dynamical flow in the  LF. At zero order, where the velocity and 
    temperature are global in a patch, the extension of the string in 1-direction
    of the boundary is given by \eqref{11} and in  the RF in the fluid we have
   \begin{equation}
   \Delta x^{1}\;=\;\frac{\pi}{4}\;b(x^{\alpha}_{0})\;u^{1}(x^{\alpha}_{0})\;=\;\frac{1}{4}\;\frac{1}{T(x^{\alpha}_{0})}u^{1}(x^{\alpha}_{0}) =\;\frac{1}{4}\;\frac{1}{T(x^{\alpha}_{0})}\gamma(x^{\alpha}_{0})\beta(x^{\alpha}_{0})\label{15'}
   \end{equation}
   where $u^{1}=\frac{\beta}{\sqrt{1-\beta^2}}$ and $x^{\alpha}_{0}$ is the
   position of the quark.
   In order to restrict string to lie in a bulk tube we must demand $\Delta x^{1}$
   not to be greater than the patch width, $1/T$.  Let us recall that since the effective temperature in a boosted thermal plasma is
 $T/\gamma$, the effective mean free path length turns out to be $\gamma/T$. Hence
 our requirement can be restated via following inequality: 
 \begin{equation}
 \Delta x^{1}<\;\frac{1}{T(x^{\alpha}_{0})}\gamma(x^{\alpha}_{0}).
 \end{equation}
As it is clear, relativistic bound on flow velocity ensures that our required condition 
is always satisfied. It simply emphasises that perturbatively computing of drag force 
in the framework of fluid/gravity is meaningful.

That the string always lies in just one bulk tube helps us to implement perturbative 
computations. The idea is to compute the effect of fluid derivatives on the  location of a special point on the world-sheet, perturbatively. At zero order this point is exactly the position of world-sheet horizon. However, as shown in \cite{Abbasi:2013mwa}, when considering the effcets of fluid gradiants, the former will no longer coincide with the world-sheet horizon.

    To compute first order corrections 
    to drag force, we make attempt to find the string solution 
    in the first-order corrected gravity background by use of our perturbative
    procedure. It will be subject of the two next sections.

  \begin{figure}
 \begin{center} 
\includegraphics[width=.45\textwidth]{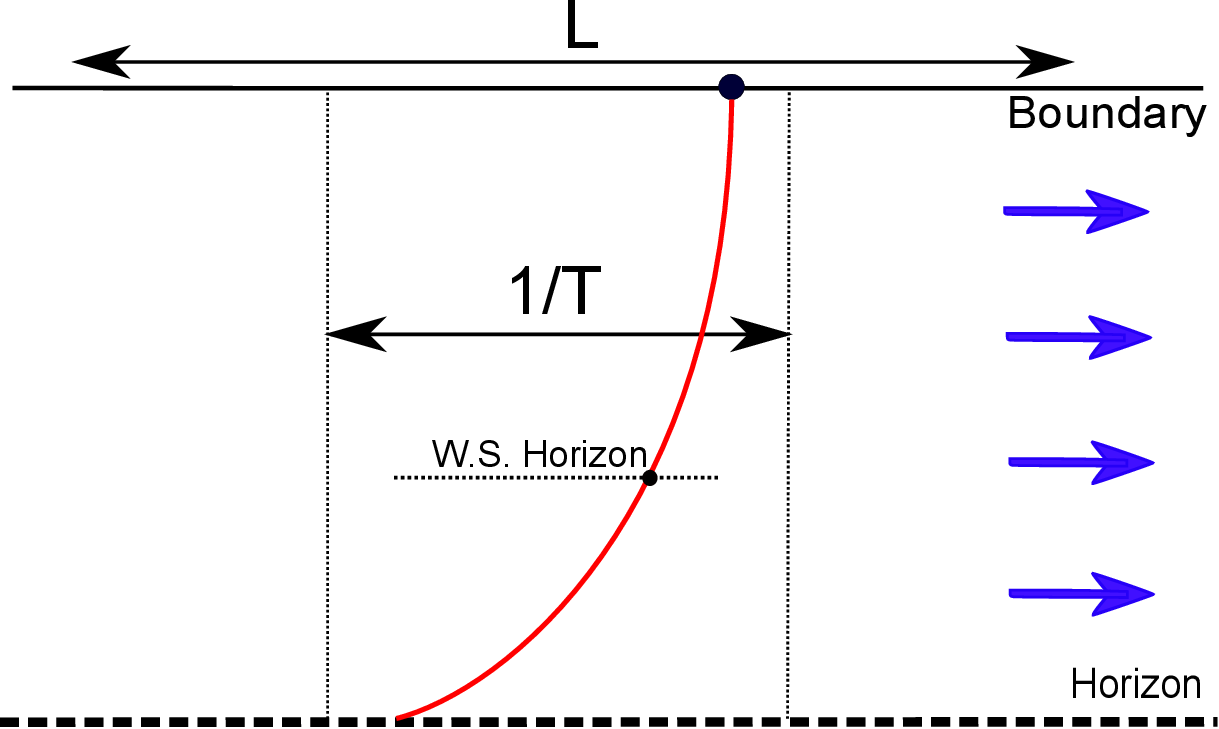}
 \caption{String in a bulk tube. In the RF and  in the 
 Eddington-finkelstein coordinates, the trail of the 
 string is in opposite direction of   the flow.}
 \label{lcdeconfined}
\end{center}\end{figure}

\section{Quark in a temperature varying plasma}
\label{5}
In order to compute derivative corrections to drag force exerted on a quark
in fluid medium, we first choose a simple but instructive example.
We will see that this toy model problem gives  more insight into
 our perturbative procedure and reveals  some of the ingredients 
discussed in \eqref{4.2}. 
 
  Let us consider a  quark uniformly moving in the
  1-direction of a globally equilibrated plasma. If the temperature 
  in the LF is globally varying  in time, it seems to be as a dynamical flow
  with  the following profile:
  \begin{equation}
  u^{\mu}_{f}=(1,0,0,0),\;\;\;\;\ T(t).  \label{16}
\end{equation}
 But as it can be  simply seen, this profile does not solve the
 free  fluid equations of motion, $\nabla_{\mu}T^{\mu\nu}=0$.
 Instead, applying an appropriate varying dilaton function,
 this profile can be accounted as a forced fluid flow on
 a flat background.  According to \cite{Bhattacharyya:2008ji}, the forced Navier-Stokes
 equations, \eqref{14}, at the level of  leading contribution
 of the dilaton function is given by:
\begin{equation}
\nabla_{\mu}T^{\mu\nu}=-(\pi T)^3\partial^{\nu}\phi\partial_{0}\phi    \label{17}
\end{equation}
where, at ideal order, the dilaton field
does not give any contribution and  therefore the temperature remains  constant in time.
 So, the given profile in \eqref{16} changes to:
\begin{equation}
 u^{\mu}_{f}=(1,0,0,0),\;\;\;\;\ T=Const. ,\;\;\;\;\; \phi(t)   \label{18}
\end{equation}
which says that at   ideal order, a rest fluid with varying temperature
 does not exist. Instead, we could have a thermal
field theory with a slowly time varying dilaton  function.
From the previous section, we  recall  that the  drag
force at   ideal order can be obtained by
localizing the thermodynamic drag  force. Restoring dilaton
factor in the computations of \eqref{3.1},  we can rewrite 
the rate of momentum loss in the fluid frame as
\begin{equation}
f^{1}=\frac{dp}{dt}=\frac{\pi}{2 \alpha'}\; e^{\frac{1}{2}\phi(t)}\;T^2\;\frac{u_{q}}{\sqrt{1-u_{q}^2}} \label{19'}
\end{equation}
  where from \cite{Bhattacharyya:2008ji}, $\Phi(t)$ is equal to $\phi(t)$ at zero order
  of gravity set up.  From the string theory we remember that
  in the presence of a nontrivial dilaton field, the string
  coupling would be changed by a dilaton factor and so the 'tHooft
  coupling \cite{Maldacena:1997re}
\begin{equation}
\lambda=\;g_{YM}^2 \;N_{c}=\;4\pi g_{s}\;N_{c}=\frac{1}{\alpha'^2}   \label{19''}
\end{equation}
 would be changed too. We see that requesting to have a varying
 temperature plasma is equivalent to varying the coupling in this
 medium. Consequently equation \eqref{19'} gives  drag force
 exerted on a moving quark in a plasma whose coupling is  slowly varying
 in time.

  Now, let us  consider a quark in this medium and
  investigate the first-derivative  corrections, namely
  viscous corrections, to drag force. According to
  the previous  section discussion, an appropriate frame to study
  the derivative corrections is the RF, so hereafter we assume the
  slowly time variation of dilaton to be occurred at  this
  frame.  At viscous order, the profile of such
  flow is given by \cite{Bhattacharyya:2008ji}
 \begin{equation}
u^{\mu}_{f}=(u^{0},u^{1},0,0),\;\;\;\;\ T(\tau) ,\;\;\;\;\; \phi(\tau)   \label{18'''}
\end{equation}
where $u^{1}=-\gamma u_{q}$ and $\tau$ is the time coordinate in the RF.
Based on  \cite{Bhattacharyya:2008ji}, the  dual picture  of such flow
in gravity side is an Einstein-dilaton system whose metric does not
 get correction up to the first order and  first-order correction to the dilaton 
 field is given by
\begin{equation}
\Phi_{cor}(x^{\mu},r)=b\;\partial_{\tau}\phi\;\int_{r b}^{\infty}\;dr\frac{r^3-1}{r^5f(r)}.  \label{19"}
\end{equation}
 As before in the static gauge $(\sigma,\tau)=(r,\tau)$, the embedding of
 the string may be written as  $X^{\mu}(r,\tau)=(r,\tau,X^{i}(r,\tau))$. 
 Taking the following ansatz
 \begin{equation}
 X^{1}=\xi(r,\tau),\;\;\;X^{2}=X^{3}=0,    \label{19}
 \end{equation}
 the  Nambu-Goto lagrangian 
will be given by

\begin{equation}
{\cal{L}} =e^{\frac{1}{2}\Phi(r,\tau)}\; \sqrt{A\xi'^{2}+ {2}B\xi'+C}.  \label{20}
\end{equation}
  In the above expression, $A=A_{1}$, $B=B_{1}+B_{2}\dot{\xi}$, $C=C_{1}+2C_{2}
  \dot{\xi}+ C_{3}\dot{\xi}^2,$ where among the  subscripted coefficients
   only $A_{1}$
  is a function of time and the others are constant (\ref{333}). Also $\xi'$
   and $\dot{\xi}$ are partial derivatives of $\xi$ with respect to $r$ and $\tau$.

     As discussed in \eqref{3.1}  we should
     find the world-sheet currents to compute the drag force. Based on our
      discussion in \eqref{2},  drag force in the present 
     case is given by \eqref{7'} similarly.
     From \eqref{20}, it's clear that  $\Pi^{\alpha}_{M}$s  are not all  conserved 
    and we have
     \begin{equation}
     \partial_{\alpha}\Pi^{\alpha}_{1,2,3}=0,\;\;\;\; \;\;\;\;\partial_{\alpha}\Pi^{\alpha}_{\tau}=f_{\tau}. \label{21}
     \end{equation}
      Note that $\alpha=r,\tau$
      are the world-sheet coordinates and $\mu=\tau,1,2,3$ are the
      boundary directions.
      The existence of time dependency in the action, does not
      allow $\Pi^{r}_{1}$ to be constant along the string.
      Physically, this variation along the string is related
      to  the energy and momentum transfer between the string
      and the non-static bulk of the space-time.  Thus by integrating
      the equation $\partial_{\alpha}\Pi^{\alpha}_{1}=0$ over  the $r$
component of the world-sheet we obtain:
\begin{equation}
\Pi^{r}_{1}(r,\tau)-\Pi^{r}_{1}(r_{0},\tau)=-\frac{d}{d\tau}\int^{r}_{r_{0}}\Pi^{\tau}_{1}(r',\tau)dr'  \label{22}
\end{equation}
where
\begin{equation}
 \Pi_{1}^r(r,\tau)=-\frac{1}{2\pi\alpha'}e^{\frac{1}{2}\Phi(r,\tau)} G_{1\nu} \frac{(\dot{X}.X)(\dot{X}^\nu)'-(\dot{X})^2(X^{\nu})'}{\sqrt{-g}}=-\frac{1}{2\pi\alpha'}\pi(r,\tau). \label{23}
\end{equation}
To study the string embedding, we derive  $\xi'$ from the above equality
\begin{equation}
 \xi'=-\frac{B}{A}\pm\frac{1}{A}\sqrt{\pi^2 e^{-\Phi(r,\tau)}\frac{B^2-AC}{\pi^2 e^{-\Phi(r,\tau)}-A}}   \label{24}
\end{equation}
with  $A$, $B$ and $C$  introduced in \eqref{20}.
Although the presence of a slowly varying dilaton function
 have changed  the numerator and denominator in comparison with
 (\ref{aa}), just like there we demand the second term in \eqref{24}
 to be real . Again, this reality condition
 obliges the numerator and denominator to have common root.
 To compute this root perturbatively,
 we add a  correction term to the embedding as
 \begin{equation}
 \xi(r,\tau)=\xi_{0}(r)+\xi_{cor}(r,\tau).   \label{24'}
 \end{equation}
  $\xi_{0}(r)$ is the zero-order embedding given in \eqref{9} in
which $u_{q}$ vanishes in RF:
\begin{equation}
\xi_{0}(r)=u^{1}_{f}\;b\;\left(\arctan(rb)-\frac{\pi}{2}\right). \label{24"}
 \end{equation}
 As a  remarkable point,  the above expression is independent of 
 $\Phi$. The other
 point is the constancy of $b$ at this order. It is a direct  consequence of
  equation \eqref{17}.
 Inserting \eqref{24'}
in \eqref{24}  and Solving $B^2-AC=0$ up to first order, we obtain the
first-order corrected root as being  $r^*(\tau)=r^{*}_{0}
=\;\frac{1}{b}\sqrt{u_{f}^{0}}$.
 We see that the  presence of such a
  dilaton function has not changed $r^*$
 up to  the  first order,  Confirming our expectation that a scalar field
 does not influences the  geometry at the  order. In order for $r^*$ to be also
 the root of the denominator
 up to the first order, we should have:
\begin{equation}
\begin{split}
\pi(r^*,\tau)&=-e^{\frac{1}{2}\Phi(r^*,\tau)}\;\frac{u_{f}^{1}}{b^2}\\
             &=-\left(1+\frac{1}{2}\;b\;u_{f}^{0}\;\partial_{\tau}{\phi}\int_{r^*_{0} b}^{\infty}
             \;dr\frac{r^3-1}{r^5f(r)}\right)e^{\frac{1}{2}\phi(\tau)}\;\frac{u_{f}^{1}}{b^2}
    \label{28}
\end{split}
\end{equation}
As clearly observed, the only viscous contribution to above expression
is related to the first-order correction of $\Phi$ given in  \eqref{19"}.

 Up to now, we have succeeded to compute the value of $\Pi_{1}^r(r,\tau)$ at
a special point on the string. Now one can simply compute
compute $\Pi_{1}^r(0,\tau)\mid_{r\to\infty}$ from \eqref{22}.  To compute
the right hand side of \eqref{22} up to the first order, the integrand
 should be worked out at zero order:
\begin{equation}
\Pi^{\tau}_{1}(r,\tau)=-\frac{1}{2\pi\alpha'}e^{\frac{1}{2}\Phi(r,\tau)} G_{1\nu} \frac{(\dot{X}.X')({X}^\nu)'-(X')^2(\dot{X}^{\nu})}{\sqrt{-g}}=\frac{1}{2\pi\alpha'}\;e^{\frac{1}{2}\Phi(r,\tau)}\;\frac{u_{f}^{0}u_{f}^{1}}{1+r^2 b^2} \label{29}
\end{equation}
Inserting  \eqref{28} and \eqref{29} in \eqref{22} with choosing
$r_{0}=r^*$, we obtain $\Pi^{r}_{1}(r,\tau)$ up to the first order as following:
\begin{equation}
\Pi^{r}_{1}(r,\tau)=\frac{1}{2 \pi \alpha'}\; e^{\frac{1}{2}\phi(\tau)}
\;\frac{u_{f}^{1}}{b^2}\;\left[1+u^{0}_{f}\;b\;\dot{\phi}(\tau)\;\left\lbrace 
\frac{1}{4}(\lim_{r\rightarrow\infty}g(r)-g(r^*_{0}))-\frac{1}{2}\left(h(r)-h(r^*_{0})\right)\right\rbrace \right] \label{30}
\end{equation}
where
\begin{equation}
\begin{split}
g(r)&=\arctan(r b)+\frac{1}{2}\log\frac{(r b)^4}{((r b)^2+1)(r b+1)^2},\\
h(r)&=\arctan(r b).   \label{30'}
\end{split}
\end{equation}
We can explicitly see that the value of $\Pi^{r}_{1}(r,\tau)$
at each point on the string is finite.
In order to obtain the rate  of momentum loss of the quark in this
medium, we rewrite \eqref{7'} but at this time
up to first order:
\begin{equation}
 F^{1}=\frac{1}{\sqrt{-g}}\Pi^{r}_{1}\vert_{r\to \infty}=\left[\left(\frac{1}{\sqrt{-g}}\right)_{(0)}\left(\Pi^{r}_{1}\right)_{(0)+(1)} +\left(\frac{1}{\sqrt{-g}}
 \right)_{(1)}\left(\Pi^{r}_{1}\right)_{(0)}\right]_{r\to \infty}.
 \label{30''}
\end{equation}
where the subscripts remark the order of perturbation.
Some tedious calculation shows that the second term in the
brackets vanishes on the boundary. So up to viscous order, the 1-component
of  drag force in the RF 
turns out to be as
\begin{equation}
F^{1}(\tau)=\frac{1}{2 \pi \alpha'}\; e^{\frac{1}{2}\phi
(\tau)}\;\frac{u_{f}^{1}}{b^2(\tau)}\;\left[1+b(\tau)\;u_{f}^{0}\;\dot{\phi}(\tau)
\;\left\lbrace\frac{1}{2}h(r^*_{0})-\frac{1}{4}g(r^*_{0})-\frac{\pi}{8}\right\rbrace\right]. \label{30'''}
\end{equation}
One can simply see from $F^{\mu}_{(RF)}=\gamma (f^{\tau},f^{1},0,0)$
that the leading contribution in the above expression is
in complete agreement with \eqref{19'}.

In analogy with \eqref{8'''}, to have  a fully covariant expression,
one can appropriately combine three vectorial objects $u^{\mu}_{q}$,
 $u^{\mu}_{f}$, $\partial^{\mu}\phi$  and construct  a general
 vector which in addition to satisfying \eqref{3'''}, is equal to
 \eqref{30'''} in the RF. Doing these stages, the result is given by
 \begin{equation}
  F^{\mu}(s_{1})=\frac{1}{2\pi \alpha'}\;\;\frac{e^{\frac{1}{2}
  \phi(s_{1})}}{b^2({s_{1}})}\;\;\left[1+u_{f}.\partial
  {\phi}\;b(s_{1}) \left\lbrace\frac{1}{2}h(s_{2})-\frac{1}{4} g(s_{2})-\frac{\pi}{8}\right\rbrace
  \right]\left((u_{q}.u_{f})u^{\mu}_{q}+u^{\mu}_{f}\right) \label{31}
    \end{equation}
where $s_{1}=x.u_{q}$ and $s_{2}=\sqrt{-u_{q}.u_{f}}$ are lorentz
scalars. Scalar nature of the dilaton says that demanding to have a
plasma with slowly varying coupling in the RF, induces  a space-time
 varying dilaton in the LF. The argument of dilaton field
in the latter case can be determined by the lorentz transformation 
of the argument in the former case:

\begin{equation}
RF:\;\;\;\phi(\tau)\;\;\;\to\;\;\;LF:\;\;\;\phi(\gamma(t-u_{q}x)).  \label{32}
\end{equation}
Such a situation seems somewhat unreal. As a well defined problem,
one may expect to be encountered with a plasma medium which its
coupling  varies in  time only. We have also answered  this problem!
One can simply observe that in the non-relativistic limit,
 $\phi$ in the LF behaves just like in the RF as a function of time.
Therefore we can say that  drag force exerted on a non-relativistic heavy quark
moving in a rest viscous fluid,  whose coupling is varying in time, is given
by:
\begin{equation}
F^{1}(t)=-\frac{1}{2 \pi \alpha'}\; e^{\frac{1}{2}\phi
(t)}\;\frac{u_q}{b^2(t)}\;\left[1+\frac{1}{16}(\frac{\pi}{2}-3\log{2})\;\dot{\phi}(t)
\;b(t)\right] \label{33}
\end{equation}

In summary, invoking the perturbative procedure explained in
\eqref{4}, we computed the viscous correction to drag force
exerted on a non-relativistic quark in a plasma medium with
a slowly time varying coupling. As it is expectable, by an 
increase in coupling of the plasma, the drag force will increase.
\section{Quark in the Bjorken flow}
\label{six}

\subsection{Geometry of the flow}
\label{geo}

Bjorken flow \cite{Bjorken:1982qr} has been known as a fine
phenomenological description of hydrodynamic regime in heavy-ion
 collisions  in last three  decades. According to Bjorken's
proposal, the post-thermalization fluid dynamics
in the central rapidity region of heavy-ion collisions 
is  longitudinal boost
invariant. This symmetry fixes the velocity  of the fluid
 automatically.  Solving the relativistic
 hydrodynamic equations of motion determines the fully profile.
 In addition, it is usually assumed  that  there is no dependence on transverse 
 coordinates. This assumption is in correspondence with the large 
 nuclei limit. 
 
 Relevant coordinates  to study Bjorken flow are the 
 $proper\;time-rapidity$ coordinates by which the Minkowski
 metric is be given by
 \begin{equation}
 \mathrm{d}s^2=-\mathrm{d}\tau^2+\tau^2\mathrm{d}\eta^2+\mathrm{d}x_{1}^2+\mathrm{d}x_{2}^2 \label{34}
 \end{equation}
 where $\tau$ and $\eta$ are proper time and space-time rapidity
 respectively
 \begin{equation}
 \tau=\sqrt{t^2-x_{3}^2},\;\;\;\;\;\eta=\frac{1}{2}\;\log\frac{t+x_{3}}{t-x_{3}}. \label{35}
\end{equation}

The ideal Bjorken  flow in this coordinates can be
described by
\begin{equation}
u^{\mu}=(1,0,0,0),\;\;\;T(\tau)\sim \frac{1}{\tau^{1/3}}. \label{36}
\end{equation}
%
\subsection{Perturbative computations}
\label{6.1}

\begin{figure}
\centering
\includegraphics[scale=.3]{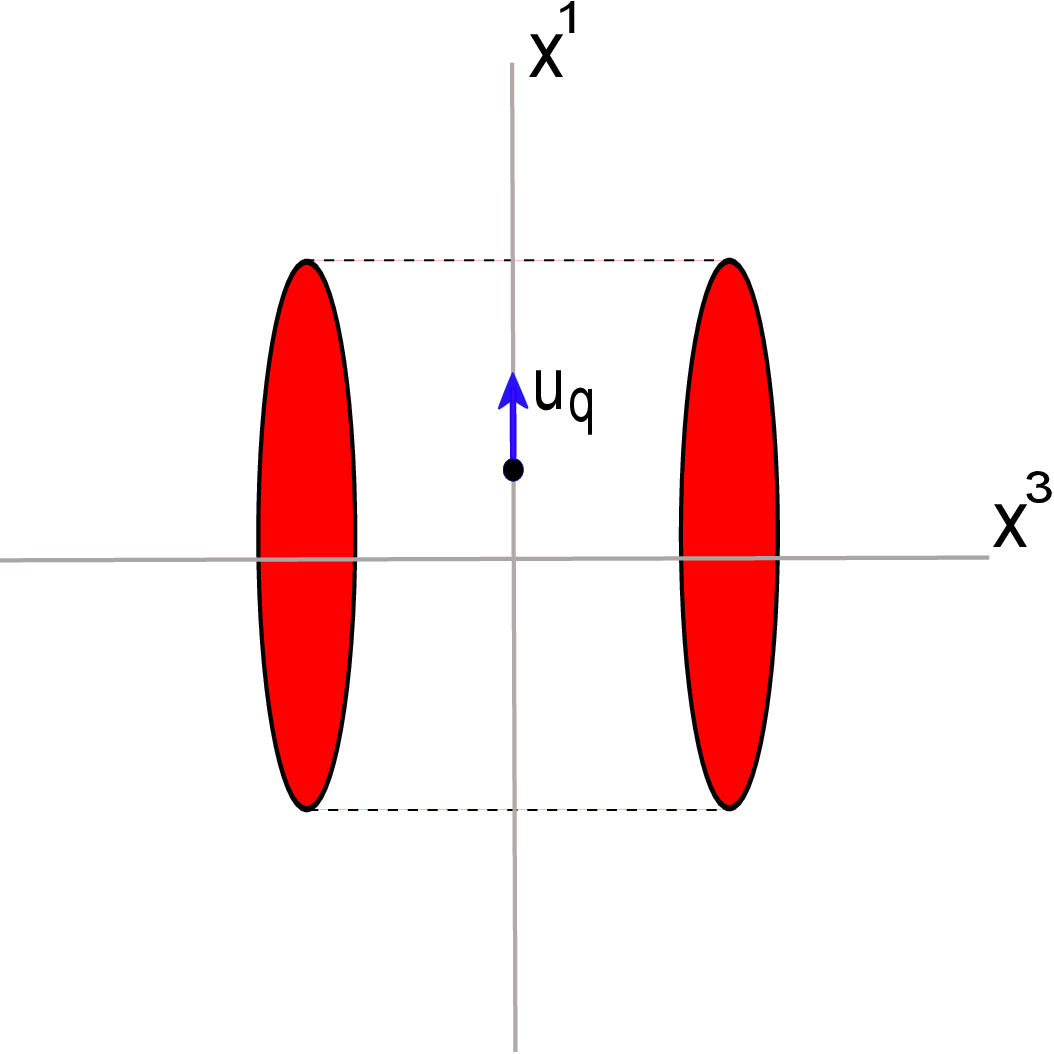}
\caption{A transverse quark in the longitudinal expanding plasma}    \label{fig:cal}
\end{figure}

At first time, the  gravity dual of  Bjorken flow
was constructed by JP in the context of AdS/CFT. In this derivation,
the bulk metric  has been computed for asymptotic states,
namely large times, of the flow by demanding the Einstein equations 
to have nonsingular  solution \cite{Janik:2005zt}. Studying  
the energy  and momentum loss of a transverse quark in 
this flow is accounted as a phenomenological problem. This issue
has been addressed in  \cite{Stoffers:2011fx,Giecold:2009wi} where
the authors have similarly applied the JP  gravity set-up to compute  
the drag force at ideal order.
We choose a different approach to compute
the drag force  up to viscous order.
In this way,  we exploit the proper time-rapidity coordinates in
 which, Bjorken flow is at rest. As depicted in Fig.\ref{fig:cal}, we restrict
the computation to the case  of a quark with zero rapidity. For such
a moving quark we have

\begin{equation}
\eta=0,\,\,\,\,\,\tau=t.    \label{36-1}
\end{equation}
%

In order to use our perturbative procedure in \eqref{4}, we make a
transverse boost to the RF.  The quark encounters with
a global boosted plasma in this frame. Symmetry considerations
allow us to get the motion in the 1-direction.  Consider the 
quark is moving with velocity $u_{q}$ in the LF. So  in the RF, 
the Bjorken flow is given by the following profile
\begin{equation}
u^{\mu}_{f}=(u^{0},u^{1},0,0),\;\;\;T(u^{0}\tilde{\tau}-u^{1}\tilde{x}^{1})\label{37}
\end{equation}
where tilded coordinates are related to the RF and $u^{1}=-\gamma_{q}u_{q} $.

The gravity dual of the above flow can be derived  from \eqref{13}
as following
\begin{equation}\begin{split}
\mathrm{d}s^2 &=  -2u_{0} \,\mathrm{d}\tilde{\tau} \mathrm{d}r-2u_{1}\, \mathrm{d}\tilde{x}^{1} \mathrm{d}r+r^2 \,(\mathrm{d} \tilde{x}^{2})^2+r^2\,\tau^2\,\mathrm{d} \tilde{\eta}^2\\&+r^2\left(\frac{u^{2}_{0}}{b^4 r^4}-1+C_{\tau \tau}\right) \mathrm{d} \tilde{\tau}^2+r^2\left(\frac{u^{2}_{0} u_{1}}{b^4 r^4}+C_{\tau 1}\right) \mathrm{d}\tilde{x}^{1}\mathrm{d} \tilde{\tau}+r^2 \left(\frac{u^{2}_{1}}{b^4 r^4}+1+C_{11}\right) (\mathrm{d} \tilde{x}^{1})^2
\label{38}
\end{split}
 \end{equation}
 where the derivative corrections are given by
 \begin{equation}
 \begin{split}
 C_{\tau\tau}&=\frac{2}{3 r}\;\frac{b\, r F(b r)\,u^{2}_{1}-u^{2}_{0}}{{u_{0}\tilde{\tau}}+u_{1}\tilde{x}^{1}}\\
 C_{\tau1}&=\frac{2}{3 r}\;\frac{u_{0}u_{1}\,(b\, r F(b r)-1)}{{u_{0}\tilde{\tau}}+u_{1}\tilde{x}^{1}}\\
C_{11}&=\frac{2}{3 r}\;\frac{b\, r F(b r)\,u^{2}_{0}-u^{2}_{1}}{{u_{0}\tilde{\tau}}+u_{1}\tilde{x}^{1}}
\end{split}
 \end{equation}
In the above expressions, for brevity, we have written $b$ for 
$b(u^{0}\tilde{\tau}-u^{1}\tilde{x}^{1})$.

 In order to construct the
 string solution in the above gravity background we choose  the
 static gauge  $(\sigma,\tau)=(r,\tilde{\tau})$. We assume that 
 \begin{equation}
 X^{1}=\xi(r,\tilde{\tau}),\;\;\;X^{2}=X^{3}=0   \label{39}
 \end{equation}
 where $X^{M}(r,\tilde{\tau})$ specifies the embedding of the 
 string in the space-time.  The Nambu-Goto lagrangian is given by
(\ref{kkk}) and  $A$, $B$, $C$ are given in \ref{444}. 
 Also $\xi'$and $\dot{\xi}$ are partial
 derivatives of $\xi$ with respect to $r$ and $\tilde{\tau}$.

There are some  noticeable differences in the drag force 
computation in comparison within  \eqref{5}. Firstly,   the
 zero-order embedding of the string is a function
of both $\tilde{x}^{1}\,$  and $\tilde{\tau}\,$  through
dependency on $b$. Secondly, since  the
translational invariance of the action has been broken in all
 boundary directions, the 
equations of motion in these directions may be written as follows
\begin{equation}
\partial_{\alpha}\Pi^{\alpha}_{\tilde{\mu}}=f_{\tilde{\mu}},\;\;\;\;\;
f_{\tilde{\mu}}=\frac{\partial{\cal{L}}}{\partial{x^{\tilde{\mu}}}}  \label{41}
\end{equation}
where $\tilde{\mu}$ are the boundary  directions in the RF. As our previous examples,
 we need to evaluate $\Pi^{r}_{1}$  at the boundary to compute the drag force. So again,
we use (\ref{aa}) and \eqref{8} by a small difference. We exchange $\pi_{\xi}$ with 
$\pi(r,\tilde{\tau})$ in this case.
Let us recall that at zero order, $B^2-AC$ vanishes
somewhere in the bulk, namely the  world-sheet horizon. Requiring the
string to be stretched from the boundary to the horizon implies that the
denominator should be also vanished at this point.  The position of this
 point will be corrected in each   order of perturbation.
 To find the first-order correction, consider
\begin{equation}
 \xi(r,\tilde{\tau})=\xi_{0}(r,\tilde{\tau})+\xi_{cor}(r,\tilde{\tau})   \label{44}
\end{equation}
where
\begin{equation}
\xi_{0}(r,\tilde{\tau})=u^{1}\;b(u^{0}\tilde{\tau})\;\left[\arctan(r\;b(u^{0}\tilde{\tau}))-\frac{\pi}{2}\right] \label{44'}
 \end{equation}
is the zero-order embedding which has been constructed
by localizing $b$ in the thermodynamic embedding in
\eqref{9}. Inserting \eqref{44} in (\ref{aa}), and
solving $B^2-AC=0$, we obtain  the first-order corrected
position the root as $r^*({\tilde{\tau}})
=r^*_{0}({\tilde{\tau}})+r^*_{1}({\tilde{\tau}})$ where
\begin{equation}
\begin{split}
r^{*}_{0}(\tilde{\tau})&=\frac{\sqrt{-u_{0}}}{b}\\
r^{*}_{1}(\tilde{\tau})&=\;\frac{1}{4\;r^{*3}_{0}}\;\left[\frac{2 u_{0} u_{z}}{b^4} \dot{\xi}_{0}(r^{*}_{0})+\;\frac{u^{0}}{b^2}G_{\tau\tau}^{(1)} (r^{*}_{(0)})\right]\,.
 \label{45}
\end{split} 
 \end{equation}

where  the argument of $b$ is $u^{0}\tilde{\tau}$ in these expressions.
 As clearly observed in \eqref{45}, $r^{*}_{0}(\tilde{\tau})$
 is the local version of \eqref{12},
 reconfirmation of our discussion in \eqref{4}. It should be noticed that
 the derivative terms have been  computed at $\tilde{x}^1=0$.
  Demanding $\pi^2(r,\tilde{\tau}) -A$
 to be vanished at the corrected $r^*$, one can simply compute the first-order
 correction to $\Pi_{1}^r(r,\tilde{\tau})$  at this point, namely $\Pi_{1}^r(r^{*},\tilde{\tau})$.
Having $\Pi_{1}^r(r^{*},\tilde{\tau})$, one can integrate the 1-component of \eqref{41}
over the $r$ to find the first-order corrected
$\Pi_{1}^r(r,\tilde{\tau})$ function. After simple but cumbersome calculations, we find
\begin{equation}
\begin{split}
\Pi_{1}^r(r,\tilde{\tau})&= \Pi_{1}^r(r^{*},\tilde{\tau})+
\int^{r}_{r^*}(f_{1}-\partial_{\tilde{\tau}}\Pi^{{\tilde{\tau}}}_{1}) \;d \mathrm{r}\\
&=\Pi_{1}^r(r^{*},\tilde{\tau})+\frac{1}{2\pi \alpha'}\; \frac{u_{1}}{b^2}
  \left(\frac{\pi}{2}-\arctan(r b)-\frac{\sqrt{u^{0}}}{1+u^{0}}\right) u_{0}\dot{b}\;.  \label{47}
\end{split}
\end{equation}
which is finite at all points on the string. In order to compute
the drag force at  viscous order, we again utilize
\eqref{30''} and obtain
\begin{equation}
F^{1}_{RF}=\frac{1}{2\pi \alpha'}\;\frac{u_{1}}{b^2}\left[1+\mathcal{A}\;u_{0} \;\dot{b}+\mathcal{B}\;u_{1}\;b'+\mathcal{C}\,\frac{b}{\tilde{\tau}}\right]
 \label{50}
\end{equation}
where
\begin{equation}
\begin{split}
\mathcal{A}&=\;\frac{1+u_0^2}{b}\,\left(\arctan(r^{*}_{0}\;b)-\frac{\pi}{2}\right),\\
\mathcal{B}&=\;\frac{1+u_0^2}{b}\,\left(\frac{\pi}{2}-\arctan(r^{*}_{0}\;b)\right),\\
\mathcal{C}&=\;\frac{1}{3b}\;\left( \frac{\sqrt{u^0}}{u^{0}}-F(br^*_{0})(\frac{1+u_{0}^2}{u^{0}})\right) \label{50-1}
\end{split}
\end{equation}
Although the finiteness of corrections is a sign of truth for
our perturbative procedure, one can separately
check it by computing the first-order corrected $\xi'$. 
  
  So far, we have computed the drag force in the RF. Because of
  breaking the rotational symmetry, we can not 
  covariantize the drag force. So in order to find the drag expression
  in the LF, we first make the inverse boost as following
 \begin{equation}
 \begin{split}
 four-velocity\,\, of\,\, the\,\, quark:\;\;\;\;\;&(1\,,0\,,0\,,0)\;\;\;\longrightarrow\;\;(\gamma\,,\gamma\, u_{q}\,,0\,,0)\\
 quark\,\, coordinates:\;\;\;\;\;&(\tilde{\tau}\,,0\,,0\,,0)\;\;\;\longrightarrow\;\;(\frac{1}{\gamma}\,t\,,u_{q}\,t\,,0\,,0).   \label{52}
\end{split}
 \end{equation} 
  It  should be reminded that for a quark with zero rapidity, the 
  proper time and the LF time coordinate are coincident. The 1-component
  of drag force in the LF may be written as
  \begin{equation}
  F^{1}_{LF}=\frac{\partial x^{1}}{\partial\tilde{x}^{1}}\;F^{1}_{RF}+
  \frac{\partial x^{1}}{\partial\tilde{\tau}}\;F^{0}_{RF}.  \label{52-1}
  \end{equation}
  For the sake of  computing the $F^{0}_{RF}$, we need to find the $\Pi^{r}_{\tau}$.
  After some calculation similar to that for obtaining the $\Pi^{r}_{1}$,
  we get that
  \begin{equation}
   F^{0}_{RF}=0    \label{52-2}  
   \end{equation}
   up to viscous order.  As we have pointed out in \eqref{2}, this condition 
   would be  maintained even when the derivative corrections are taken into account.
   Physically, this means that there should not be any energy loss 
   in the RF at viscous order, just like at ideal order. Equation \eqref{52-2} is 
   in complete agreement with this constraint. So substituting \eqref{50} in 
   \eqref{52-1} gives the rate of momentum loss in the LF.  
   But to proceed further, it is needed to have the  relations between the 
   derivatives in   these two frames which can be written as
  \begin{equation}
  \begin{split}
  b'_{RF}(u^{0}\tilde{\tau})&=\gamma\,u_{q}\;\dot{b}_{LF}=\;-\;u^{1}\;\dot{b}_{LF}\\
  \dot{b}_{RF}(u^{0}\tilde{\tau})&=\gamma\;\dot{b}_{LF}(t)=\;u^{0}\;\dot{b}_{LF}    \label{53}
  \end{split}
  \end{equation}
  where $u^{0}=\gamma$ and $u^{1}=-\gamma u_{q}$. Finally the rates of 
  energy and momentum loss of the quark in the LF turn out to be as 
  \begin{equation}
  \frac{dp}{dt}=\frac{1}{u_{q}}\frac{dE}{dt}=\frac{-1}{2\pi\alpha'}\frac{u_{q}\gamma}{b^{2}}\left[1-\dot{b}\;(\gamma^2 \,\mathcal{A}+\gamma^2\, 
  u_{q}^2\,\mathcal{B})+\frac{b}{t}\;\gamma\,\mathcal{D}\right] \label{54}
  \end{equation}
  where 
  \begin{equation}
  \begin{split}
  \mathcal{A}&=\;-\mathcal{B}=\;\frac{1+\gamma^2}{b}\,\left(\arctan(\sqrt{\gamma})-\frac{\pi}{2}\right),\\
    \mathcal{D}&=\frac{1}{3b \gamma}\;\left( \sqrt{\gamma}-\frac{}{}
    F(\sqrt{\gamma})(1+\gamma^2)\right). \label{55}
  \end{split}   
  \end{equation}
 In the expression (\ref{54}), one can simply show that $(\gamma^2 \,\mathcal{A}+\gamma^2\, 
  u_{q}^2\,\mathcal{B})=\mathcal{A}$. 
  
  As a result, by use of $\dot{b}=\frac{b}{3 \tau}$ at $\eta=0$, we can write the first order corerected drag force exeted on quarks moving through the Bjorken flow on the path $x^3=0$ as being:
   \begin{equation}
   \boxed{
 f_{\text{drag}}= f_0+ f_1=\frac{dp}{dt}=\frac{1}{u_{q}}\frac{dE}{dt}=\frac{-1}{2\pi\alpha'}\frac{u_{q}\gamma}{b^{2}}\left(1+ c(\gamma) \frac{b(t)}{3t}\right)} \label{54}
   \end{equation}
  with
  \begin{equation}
  c(\gamma)=b(t)\big(3 \gamma \mathcal{D}-\mathcal{A}\big)=\,\sqrt{\gamma}+(1+\gamma^2)\left(
  \frac{\pi}{2}-\arctan(\gamma)-F(\gamma) \right)
  \end{equation}
  Formula (\ref{54}) may be also written in terms of temperature and its gradiant:
    \begin{equation}
 f_{\text{drag}}= f_0+ f_1=\,-\frac{\pi}{2}\sqrt{\lambda}\frac{u_{q}}{\sqrt{1-u_{q}^2}}T^2(t)
  \left(1+ c(\gamma) \frac{1}{3\pi T(t) t}\right). \label{56}
  \end{equation}
According to our expectation, drag force at ideal order is
the same as the drag force in thermal plasma medium but 
with a difference that the temperature would be localized.
  
  In \cite{Lekaveckas:2013lha}, the drag force exerted on a quark moving on a straight path in Bjorken flow has been computed (see formula (4.5) in \cite{Lekaveckas:2013lha}). In the notation of \cite{Lekaveckas:2013lha} when $z=0$, $\gamma_v=0$, namely the quark is moving in $\eta=0$ plane. The drag force is  
    \begin{equation}
    	\vec{f}(\tau)=\,-\frac{\sqrt{\lambda}}{2\pi\, b(\tau)^2}\gamma\,\left(1+\alpha_{2}\left(\gamma\right)\frac{b(\tau)}{3 \tau} \right)\, 
    	\left( 
    	{\begin{array}{c}
    		\beta_{x}\\
    		\beta_{y}\\
    		0\\ 
    		\end{array} } 
    	\right)
    \end{equation}\label{finalBjorken}
  whee $\alpha_2(x)=c(x)$. This formula is clearly in full agreement with the formula (\ref{54}) given above.
  
  At the end, we should check the finiteness of the string 
  extension in the 1-direction at viscous order. One can 
  accomplish this lengthy and cumbersome calculation and confirm that 
  $\xi(r)$ has finite range at first order. However, the
  finiteness of the drag corrections is the implicit confirmation of perturbative method.

  Now we are ready to use our perturbative results to obtain 
estimates on the  contribution of the drag force  on a 
transverse quark in expanding QGP.   To proceed, we consider 
the drag coefficient $\eta_{D}$ which is defined as follows:
\begin{equation}
\eta_{D}=-\frac{1}{p}\,\frac{dp}{dt}   \label{59}
\end{equation}
  and to find it in our case, we rewrite the drag force expression 
  in \eqref{56} as a function of the quark momentum:
  \begin{equation}
   \frac{dp}{dt}=-\frac{\pi}{2}\sqrt{\lambda}\frac{p}{M}T^2(t)
  \left[1+\frac{\dot{T}(t)}{\pi T^2(t)}(\frac{p^2}{M^2}\,
  \mathcal{B}+(1+\frac{p^2}{M^2}) \,\mathcal{A})+\frac{1}{\pi T(t) t}
  (1+\frac{p^2}{M^2})\,\mathcal{D}\right]. \label{60}
  \end{equation}
  $\mathcal{A}$, $\mathcal{B}$ and  $\mathcal{D}$ have been introduced in \eqref{55}
  which should be expressed in terms of $p/M$ in the above equation. In contrast
  to the constant drag coefficient of a quark moving in thermal plasma,
  the drag coefficient is a function of the quark momentum here. 
  Although this feature appears only in viscous order, there is no a constant
  $\eta_{D}$ even at ideal order. the time-dependency of $T$ leads to 
  variation of the drag coefficient.

\section{Discussion}
\label{discuss}
\label{7}
Applying AdS/CFT correspondence, several problems related
to a moving quark  in thermal plasma have been studied
in resent years\cite{Herzog:2006gh,Gubser:2006bz,Akhavan:2008ep, CasalderreySolana:2009ch,
CasalderreySolana:2009ch,Fadafan:2008bq,Chernicoff:2008sa,AliAkbari:2011ue,Liu:2006ug,deBoer:2008gu,
Ebrahim:2010ra,Kiritsis:2011ha} and etc.
In all these problems, the solutions of  classical string in gravity side,
have been strong enough to address and clarify some questions in the strongly
coupled CFT side.

It has been accepted that the expanding plasma medium generated 
after  scattering of two nuclei evolves from a far-from-equilibrium
state to the hydrodynamic regime during about 1fm. So to be more close
to the real experiment, one may be interested in studying the problems mentioned 
in  last paragraph in a fluid dynamical flow.  To proceed, a string probe
has to be studied in a gravity background dual to the boundary 
fluid dynamical flow.

In this paper, we showed that based on the fluid/gravity
correspondence, one can construct a perturbative procedure
to find the string solution on a perturbed AdS background dual
to a fluid dynamical flow on the boundary of AdS. Throughout computing the drag force
 we understood that in Eddington-Finkelstein coordinates the
string never extends beyond a bulk tube. That why such situation is always met is 
a little bit strange in gravity side. However in the  boundary theory it has a simple interpretation:
drag force must be a local function through the fluid variables.

From the phenomenological point of view, our perturbative method might make
improvement  in  previous estimates of quark diffusion time. Our result
show that in contrast to a thermal plasma, the drag coefficient changes to 
be a time dependent function in a dynamical flow.  Accordingly, the diffusion 
time will be a function of the quark creation time.

  This method can open a window to some new problems. As an interesting problem one 
  can try to compute energy loss in an anisotropic fluid, generalizing \cite{Fadafan:2012qu}. However firstly, the gravity background
  dual to an anisotropic boundary flow has to be specified.
  
   The most important problem that we wish  to answer is computing the drag force corrections in a general fluid
   dynamical flow. If it is desired, we will be able to study the effect of viscous corrections on the motion of a general 
   quark in the Bjorken flow, not only quarks restricted to move in zero rapidity plane.
   We leave the study on the issue to our further work.
\subsection*{Acknowledgements}
We would like  to thank    A. Akhavan, M. Ali-Akbari, F. Ardalan, K. Bitaghsir Fadafan, H. Ebrahim,
 A.E. Mosaffa,  F. Omidi,  N. Sadoughi,  H. Soltanpanahi,  for discussions.  We are grateful to  
M. Alishahiha, D. Allahbakhsi  for many helpful  discussions  and  comment on the manuscript. The authors wish to express their deep appreciation to H. Arfaei who was generous enough to share his expertise with them and reading the article through. N.A. would also like to especially thanks to Prof. H.Arfaei
for his  encouragements and supports. A.D. has been supported by Bonyade Melli Nokhbegan.

\appendix
\section{Appendices}
\label{Appendix}

\subsection{Global boosted plasma}\label{111}

In this appendix we rewrite the induced metric on the world-sheet given in 
\eqref{4} with more detail.  From \eqref{4} we have 
\begin{equation}
g=-(A\xi'^{2}+ {2}B\xi'+C)
\end{equation}
in which, the coefficients constructed from metric \eqref{4'} are given by
\begin{equation}
\begin{split}
A&=\;G_{ t1}^{2}-G_{11}G_{tt}\;=\;r^4\; f(br)\\
B&=\;G_{tr}G_{t1}-G_{tt}G_{r1}+u_{q}\;(G_{rt}G_{11}-G_{r1}G_{t1})\;=\;\gamma\; r^2\;(u_{q}-\beta)\\
C&=\;(G_{rt}-u_{q}\;G_{r1})^2\; =\; \gamma^2\;(1-\beta\;u_{q})^2
\end{split}
\end{equation}
where $u_{q}$ and $\beta$ are the three-velocity of quark and plasma in the 
LF respectively.

\subsection{Fluid dynamical correction} \label{222}
Our perturbative computations need to have first-order corrected dual
gravity. So in following, we recall the computed corrections in 
\cite{Bhattacharyya:2008jc,Bhattacharyya:2008ji} up to first order.

\subsubsection{Free fluid}  \label{2221}
The first-order correction to \eqref{13} has been given in equation (4.24)
of \cite{Bhattacharyya:2008jc} as follows:
\begin{equation}
\mathrm{d}s^2_{cor}= r^2 \,b\, F(b\, r)\, \sigma_{\mu\nu}  \, dx^{\mu} dx^{\nu} 
+{2\over 3} \, r \, u_{\mu}u_{\nu} \,\partial_{\lambda} u^{\lambda} \, dx^{\mu}dx^{\nu} -  r\, u^{\lambda}\partial_{\lambda}\left(u_\nu u_{\mu}\right)\, dx^{\mu} dx^{\nu}   \label{100}
\end{equation}
where
\begin{equation}
F(r) =\;{1\over 4}\, \left[\ln\left(\frac{(1+r)^2(1+r^2)}{r^4}\right) - 2\,\arctan(r) +\pi\right] 
\end{equation}
and 
\begin{equation}
\sigma^{\mu\nu}= P^{\mu \alpha} P^{\nu \beta} \, 
\, \partial_{(\alpha} u_{\beta)}
-\frac{1}{3} \, P^{\mu \nu} \, \partial_\alpha u^\alpha. 
\end{equation}
%
\subsubsection{Forced fluid} \label{2222}
In this case, the correction to the  metric has  been given 
in equation (3.4) of  \cite{Bhattacharyya:2008ji} which is 
similar to the \eqref{100} with this difference that the derivatives
are covariant derivatives. The first-order correction to the dilaton
field is given by
\begin{equation}
\Phi_{cor}=b\;u.\partial{\phi}\;\int^{\infty}_{rb}\;dx\frac{x^3-1}{x^5f(x)}.
\end{equation}
%

\subsection{Time varying plasma} \label{333}
In this appendix we rewrite all coefficient introduced in
\eqref{20} with their expressions. Note that  in following, 
we have applied first-order corrected metric in \eqref{155}:
\begin{equation}
A_{1}=\;G_{ \tau1}^{2}-G_{11}G_{\tau\tau}\;=\;r^4\;f(br),
\end{equation}
\begin{equation}
\begin{split}
B_{1}&=\;G_{r\tau}G_{1\tau}-G_{r1}G_{\tau\tau}\;=-u_{f}^{1}\;r^2\\
B_{2}&=\;G_{r\tau}G_{11}-G_{r1}G_{\tau 1}\;=u_{f}^{0}\;r^2
\end{split}
\end{equation}
and
\begin{equation}
\begin{split}
C_{1}&=\;G_{r\tau}^2\;=(u^{0}_{f})^2\\
C_{2}&=\;G_{r\tau}G_{r1}\;=-u^0_{f} u^1_{f}\\
C_{3}&=\;G_{r1}^2\;=(u^{1}_{f})^2.
\end{split}
\end{equation}
%
\subsection{Bjorken flow} \label{444}
The coefficient introduced below \eqref{39} have the same 
expressions as which given in previous appendix by this 
difference that inhere our $\tau$  should be 
substituted by $\tilde{\tau}$.

\bibliographystyle{utphys}

\providecommand{\href}[2]{#2}\begingroup\raggedright\endgroup

\end{document}